\documentclass[conference]{IEEEtran}

\usepackage[T1]{fontenc}
\usepackage{url} 
\usepackage{pifont}
\usepackage{todonotes}
\usepackage{quoting}
\usepackage{graphicx}
\usepackage{listings}
\usepackage{booktabs}
\usepackage{algorithm}
\usepackage{algorithmic}
\usepackage{tikzscale}
\usepackage{filecontents}
\usepackage{tikz}
\usepackage{subcaption}
\usepackage{hyperref}
\usetikzlibrary{decorations,arrows,shapes}
\usepackage[most]{tcolorbox}
\usepackage{color, colortbl}
\usepackage[export]{adjustbox}
\usepackage{tcolorbox}
\usepackage{tikz,pgfplots}

\definecolor{lightgreen}{rgb}{0.56, 0.93, 0.56}
\definecolor{moonstoneblue}{rgb}{0.45, 0.66, 0.76}
\definecolor{beige}{rgb}{0.9, 0.8, 0.5}
\definecolor{darkgreen}{rgb}{0, 0.55, 0.45}
\definecolor{backquote}{RGB}{221, 181, 118}
\definecolor{mGreen}{rgb}{0,0.6,0}
\definecolor{mGray}{rgb}{0.5,0.5,0.5}
\definecolor{mPurple}{rgb}{0.58,0,0.82}
\definecolor{backgroundColour}{rgb}{0.95,0.95,0.92}
\definecolor{javared}{rgb}{0.6,0,0} 
\definecolor{javagreen}{rgb}{0.25,0.5,0.35} 
\definecolor{javapurple}{rgb}{0.5,0,0.35} 
\definecolor{javadocblue}{rgb}{0.25,0.35,0.75} 
\definecolor{mygray}{RGB}{220,220,220} 

\definecolor{mygreen}{RGB}{0, 100, 0}
\definecolor{myred}{RGB}{204, 0, 0}

\definecolor{fillgrey}{rgb}{0.83,0.83,0.83}
\tikzstyle{node}=[
draw=black,
rectangle,
fill=fillgrey,
text centered
]

\lstdefinestyle{CStyle}{
	backgroundcolor=\color{backgroundColour}, 
	basicstyle=\scriptsize\ttfamily,  
	commentstyle=\color{mGreen},
	keywordstyle=\color{magenta},
	numberstyle=\tiny\color{mGray},
	stringstyle=\color{mPurple},
	breakatwhitespace=false,         
	breaklines=true,                 
	captionpos=b,                    
	keepspaces=true,                 
	numbers=left,                    
	numbersep=5pt,                  
	showspaces=false,                
	showstringspaces=false,
	showtabs=false,                  
	tabsize=2,
	language=C
}

\pgfplotsset{
	legend image with text/.style={
		legend image code/.code={%
			\node[anchor=center] at (0.3cm,0cm) {#1};
		}
	},
}

\def\inline{\lstinline[basicstyle=\ttfamily]}

\newcommand{\toolname}{{\bf Fex}}

\newtcolorbox{colorquote}[1][]{%
    enhanced, breakable, 
    size=small,
    frame hidden, boxrule=0pt,
    sharp corners,
    colback=backquote!10,
    borderline horizontal={.5pt}{0pt}{backquote!30},
    #1
}

\title{\toolname{}: Assisted Identification of Domain Features from C Programs}
\begin{document}
\author{\IEEEauthorblockN{Patrick Müller\IEEEauthorrefmark{1},
Krishna Narasimhan\IEEEauthorrefmark{1},
and Mira Mezini\IEEEauthorrefmark{1}}
\IEEEauthorblockA{\emph{mueller,  kri.nara and mezini (@cs.tu-darmstadt.de)}}
\IEEEauthorblockA{\IEEEauthorrefmark{1} TU Darmstadt, Germany}}%
\maketitle	
\begin{abstract}
Modern software typically performs more than one functionality. These functionalities or features are not always organized in a way for modules representing these features to be used individually. Many software engineering approaches like programming language constructs, or product line visualization techniques have been proposed to organize projects as modules. Unfortunately, much legacy software suffer from years or decades of improper coding practices that leave the modules in the code almost undetectable. In such scenarios, a desirable requirement is to identify modules representing different features to be extracted. In this paper, we propose a novel approach that combines information retrieval and program analysis approaches to allow domain experts to identify slices of the program that represent modules using natural language search terms. We evaluate our approach by building a proof of concept tool in C, and extract modules from open source projects. 
\end{abstract}
\section{Introduction}
\label{sec:introduction}
Large-scale software projects typically consist of a composition of domain features, which often also come in different variants. For instance, an embedded software system that manages a modern automobile is typically composed of 
 features such as the multimedia system, tyre pressure monitors, etc. 
 To manage the complexity of development and evolution of such software, it is desirable that domain features are modularized in well-defined code modules and their dependencies and variants are systematically modelled in product-lines~\cite{Clements2001}.

Unfortunately, when the need to manage or restructure existing software into a product line arises, it is often 
very challenging to do so, due to one of the following reasons: (a) the system has been developed without 
a product line in mind, or (b) inappropriate language constructs are used to structure the program into modules.  
In case (a), we need to identify code that pertains to the implementation of individual domain features.
Code that implements a single feature can be spread across multiple 
files inside the project.\footnote{In the rest of the paper, we will sometimes use the word \texttt{module} to refer to code scattered throughout the project that represent a feature.}
Berger et al.~\cite{berger:splc2015} conducted an industrial study on the significance 
and difficulties of identifying features in large-scale projects. 
In case (b), we refer to preprocessor directives like \texttt{\#ifdef} that
the C programming language offers to modularize software along the 
domain features. The C preprocessor has been found to be notoriously harmful owing to
obtrusive syntax and hindrance to comprehension and 
maintainability~\cite{malaquias:icpc2017}. 
Some approaches have been proposed 
to improve comprehension of projects organized using preprocessors, most notably the work from Le et al.~\cite{le:vlhc2011} or \texttt{PEoPL}~\cite{behringer:icse2017}.
But, they require the code to be well-structured using \texttt{ifdef}s in the first place. 
Moreover, given that most modern languages have done away with the preprocessor pragmas, 
we need a better solution that avoids preprocessors altogether.

In this paper, we propose a novel approach to identifying code that pertains 
to the implementation of a domain feature of interest, as a prerequisite to
re-engineering legacy software into first-class feature modules 
and compositions thereof.
Program slicing is a common static program analysis technique 
to extract sections of code that depend on a particular piece of data (variables in code). 
But, domain experts may not be able to determine variables to serve as starting points 
for extracting a module. 
To bridge this gap, we develop a program analysis technique that synergistically combines 
information retrieval (IR) with control- and data-flow analyses, 
thus yielding what we call a \emph{natural 
code analysis} approach. Based on this analysis infrastructure,
we implement a slicing technique that takes natural language 
terms as input. We use the term \texttt{slicing criteria} to represent the input provided by the user, which 
in our approach is comprised of the feature anchor and a similarity threshold. The former serves as input search criteria which the user provides as a natural language term. 
The latter is a value between 0 and 1 that determines how close the terms in the program should be to the input term. 
One can think of the similarity threshold as a slider that the domain expert can control to arrive at the desired module. 

More specifically, our information retrieval  
creates a program \texttt{Corpus} 
- an index that maps terms  to locations in the program - and 
internally maintains a map of the corpus to the control flow graph (CFG) of the program. 
We slice the corpus based on the slicing criteria and use the map to identify nodes in the CFG 
that correspond to this corpus slice. 
Given the identified CFG nodes, we apply data-flow analysis to gather their control and 
data dependencies. The result is a module that contains the code 
that corresponds to the input term -- representing a domain feature. 
Our approach is not intended to replace manual effort, i.e., to be fully automatic. On the contrary, we believe 
that iterative input from domain experts is a crucial aspect to the module extraction process. 
The goal is to assist the process by relieving the human expert from the heavy lifting.

Program analysis and slicing techniques are generally driven by data or program variables. To the best of our knowledge, this is the first approach that performs feature-anchored program analysis by embedding information retrieval into domain-driven program slicing.

To understand better where such an feature identification approach could be useful as a first step to re-engineer modules, consider the following scenarios:
\begin{itemize}
\item \textbf{Grbl:}~\footnote{\url{https://github.com/gnea/grbl}} is a high-performance controller for CNC milling, a cutting tool that is mounted on a rotating spindle to selectively remove material from a dedicated workpiece. The axis module, that is concerned with the control of motors inside the mill are currently deeply tied into an ocean of C-code making it hard to re-use or analyse independently. Our running example in this paper is a simplified snippet of the grbl code illustrating how our approach can identify the subset of the controller concerned with axis-related tasks.
\item \textbf{Language transpiling: } is the task of transforming sections of a software from one language into another. Such transpilations are useful with the advent of safe languages like Rust which serve as an alternative to unsafe languages like C. Companies like Mozilla are currently transpiling their C-based projects into Rust. But they do this incrementally on a feature-level to avoid a big-bang transpiling that could consume too much time and human resources. An approach such as ours could help alleviate the pain of identifying the pieces of code concerned with the feature that needs to be transpiled. As part of our evaluation, we have extracted a module from a popular grep alternative called Silver-Searcher, which is equivalent to a module transpiled into Rust and available as a pull request in the project upstream. 
\item \textbf{Linux GKI: } or the Generic Kernel Image is a proposal from the Linux core team to create a single kernel image that will be used in all devices and providers to encourage re-use. Currently, there exist multiple provider-specific kernels. In order for such a project to manifest, providers will have to identify and extract code specific to their unique features that have been now merged into their provider-specific kernels. An approach such as ours could alleviate the pain involved in such laborious identification of features. 
\end{itemize}

We evaluated our approach on six open source C projects: {\textbf{Parson, inotify, fping,}  \textbf{silver-searcher, redis}} and \textbf{memcached}, ranging from $3k$ to $165k$ lines of code, from which we extracted eight different modules in total.
We apply the tool that implements our approach, \toolname{} to these benchmarks to extract modules 
that represent significant features, e.g., the module that parses json files in the Parson project.
We compare the results against a set of modules that we extracted manually based on 
detailed analysis and understanding of the code-base to serve as a ground truth. 

We put several measures in place to make sure that manually extracted modules 
represent meaningful program features. To start with, the extracted modules do 
not break existing functionality -- existing test-cases pass.

Second,

two of the authors with three respectively four years of experience independently agreed that the manually extracted modules represent program features. Third, one of our extracted modules is externally validated because it corresponds to a module for which there is an ongoing pull request for transpiling it from C into Rust. We compare the results of \toolname{} with an extraction using \textit{grep}.

Our results show that the extracted modules 

are close to the ground truth modules in terms of matching lines of code. 

For deviating lines of code, we perform a thorough manual analysis to understand the root causes and to provide insights on how the gap could be closed. As the application of our tool on \textbf{redis}  -- the largest benchmark in our set ($165k$ LoC) -- shows, the approach scales well to code written for real-world software.

\noindent
To sum up, the contributions in this paper are as follows:
\begin{itemize}
\item 
A novel information retrieval (IR) approach to work on C files like text documents so that they can be queried using natural language. Given a C program (a set of C files) one can use the IR technique to create a corpus of the project, which is persisted for subsequently performing queries based on terms for re-engineering purposes. 
\item A novel program slicing technique for C that makes use of natural language terms as slicing criteria and uses the corpus built by the IR technique to identify the code elements pertaining to a slice. 
\item A tool that implements our approach, \toolname{} which we use to extract modules from open-source C projects with very encouraging results.

\end{itemize}

The remainder of the paper is organized as follows. 

We present our methodology in Section~\ref{sec:approach} and our evaluation experiments and results in Section~\ref{sec:eval}. We discuss threats to validity and outline future work in Section~\ref{sec:threats}, related work in Section~\ref{sec:relatedwork} and conclude in Section~\ref{sec:conclusion}.

\section{Approach}
\label{sec:approach}
\begin{figure}
\includegraphics[width=0.5\textwidth]{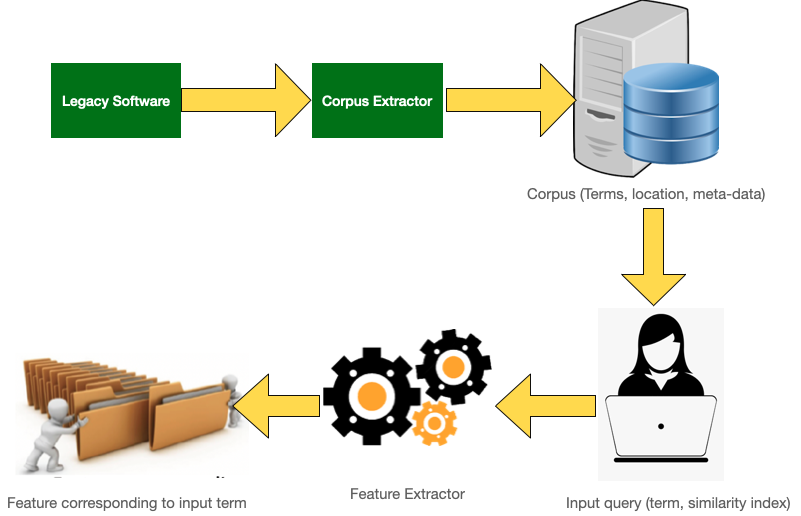}
\caption{Approach Overview}
\label{fig:approverview}
\end{figure}

An overview of the workflow supported by our approach is depicted in Figure~\ref{fig:approverview}. Given some legacy software, the extractor creates the project corpus (cf. Section~\ref{sec:corpuscreation}), which is stored in a database to ensure that the extractor is not required to be re-run for every new query.
Once the corpus is ready, the user is prompted for a keyword 
representing the desired feature in domain terms, 
which is used to filter the corpus (Section~\ref{sec:corpusslice}). 
Finally, the feature extractor maps the filtered slice of the corpus to 
syntactically correct code that represents the implementation of the
desired feature (Section~\ref{sec:extrfeature}).

\subsection{Creating the corpus}
\label{sec:corpuscreation}

The creation of the corpus is the responsibility of the \textbf{Corpus Extractor}.
Its input is a C project with a set of source files - the output is the \texttt{corpus} for the input project.

The corpus is a matrix that maps terms occurring in code to the following attributes of them extracted from the code:

\begin{itemize}
\item \textbf{Locations} where the term occurs (line, column)
\item \textbf{Program context} in which the term occurs (comment, preprocessor macro, identifier, function definition)
\item \textbf{Weight} - determines how important a term is within each function in code. There is one weight entry for each pair of function and term. Hence, weight is in fact a matrix itself - the Term Document Matrix (TDM). There are different strategies to determine the weight of a term in a function - the most simple one is to count the number of occurrences a given term has in a function. 
\end{itemize}

A \textbf{Term document matrix (TDM)}~\cite{ZHAO2013105} is a matrix that represents how frequently a term occurs in a document. The rows of the matrix represent individual documents and the columns represent the terms. There are various methods to decide what values are represented in the matrix, the most obvious one being to encode the number of occurrences of each term.  Consider the two documents:
\begin{enumerate}
\item D1 = "Time heals everything"
\item D2 = "Time cures everything"
\end{enumerate}
The TDM for the above document would be:
\begin{table}[H]
\begin{tabular}{|l|l|l|l|l|}
\hline
\textbf{}   & \textbf{Time} & \textbf{heals} & \textbf{cures} & \textbf{everything} \\ \hline
\textbf{D1} & 1             & 1              & 0              & 1                   \\ \hline
\textbf{D2} & 1             & 0              & 1              & 1                   \\ \hline
\end{tabular}
\end{table}

In natural language processing, a popular alternative to the "number of occurrences" to describe the frequency is to use the \textbf{Term frequency–inverse document frequency}\footnote{\url{https://nlp.stanford.edu/IR-book/html/htmledition/queries-as-vectors-1.html}} or in short \textbf{TFIDF}, which we use in our corpus as \textbf{weight}. This is a statistic that represents the importance of a term in a document. This value is directly proportional to the number of times a term occurs in a document, and the number of documents that contain this term.
In order to query for a term you have to create a query vector, which has 1 at the index that corresponds to the input term and 0 in all other indices, e.g., 
$\begin{pmatrix}
	\ldots &0 &1&  0& \ldots
\end{pmatrix}$. 

As one can imagine, the size of the TDM can get quite large for a big project with lots of terms. For this purpose, \textbf{Latent Semantic Indexing (LSI)}~\cite{deerwester-indexing-1990} is an indexing and retrieval technique that groups terms that are semantically equivalent into concepts, and finds the connections between these terms and concepts. LSI then uses a matrix decomposition technique called \textbf{Single value decomposition (SVD)}~\cite{trefethen97} to reduce the size of the matrix based on the relationship between the terms and the concepts.

To create the corpus for the given project, the extractor performs the following steps:
\begin{enumerate}
	\item \textbf{Tokenize the source code}: 
	We use a lexer to tokenize the source files. This will allow us to manage the individual words in our project.
	
	\item \textbf{Produce the documents}: We remind the reader that our information retrieval process requires the source files to be available as documents. To enable retrieving information at a fine granularity level, we map source files into several documents, rather than having one document per file. Documents represent one of several different constructs in the original source code, i.e., a single function is a document, and all declarations in one source file are a document on their own. 
	
	\item \textbf{Normalize the tokens}: We perform a post-processing on the tokens to normalize them, which results in \emph{terms}. During the normalization we transform all tokens to lower case and split them for different naming styles, e.g., snake case or camel case, while also retaining the original version. For example, the token  \emph{(parse\_axisCommand)} gets normalized to the terms (parse, axis, command, axisCommand, and parse\_axiscommand). In addition, we retain the program context for all term occurrence locations, i.e., whether the term is an identifier, is guarded by a preprocessor macro, or occurs within a comment. 
	
	\item \textbf{Shorten the corpus}: We then reduce the size of the resulting corpus by filtering out C keywords.
	
	\item \textbf{Create the TDM}: As already mentioned, the TDM maps pairs of documents and terms, $(d,t)$ to a number $n$. What $n$ represents depends on the chosen weighting scheme. In our default case of TFIDF, $n$ corresponds to the relative importance of $t$ inside the document $d$. For the sake of simplicity, in Table~\ref{fig:corpus}, we show the TDM for a  single function (i.e., the TDM is reduced to a single column \texttt{Weight})

	\item \textbf{Reducing the matrix}: For big projects with many functions, the corpus can become quite large. Latent Semantic indexing~\footnote{\url{https://nlp.stanford.edu/IR-book/html/htmledition/latent-semantic-indexing-1.html}} 
	
	uses a matrix factorization technique called Single value decomposition (SVD) to reduce the size of a matrix. The resulting matrix exists in what is called the SVD space. LSI also takes care of mapping the input query on the full TDM to the SVD space.

\end{enumerate}

For illustration, consider the code in Figure~\ref{fig:code} and its corresponding corpus in Table~\ref{fig:corpus}. In contrast to a TDM, this shows the corpus information corresponding to a term in a single row, i.e., a transposed TDM, and occupies only one column, because our example code contains only one function (i.e., there is only one document). In a typical case, the TDM column will have multiple sub-columns, one for each document, indicating whether it contains the term. The most frequently occurring term in the example is \textbf{command}, which has the highest TDM entry value and the least occurring terms have a TDM entry value of \textbf{0.30}.

\begin{figure}
\begin{lstlisting}[language=C++,numbers=left]
void parse_command(char* input) {
  axis_command = NULL;
  if (input[0] == 'G') {
    // mm or inches
    unit = parse_unit(input); 
    axis_command = 
      parse_axis_command(input);
    mode = 0;
    move_x(axis_command);
    move_y(axis_command);
  } else if (input[0] == 'H'){
    coolant();
  } else { FAIL(UNSUPPORTED_COMMAND);  }
  if (axis_command) {
    do_command(mode);
  }
  return;
}
\end{lstlisting}
\caption{Example Code}
	\label{fig:code}
\end{figure}

\begin{table}
	\captionsetup{singlelinecheck=off}
	\caption[test]{Corpus for running example, Locations are tuples of (Line $\times$ Column $\times$ Context), where Context is\\ \textit{i} for Identifier, \textit{c} for Comment and \textit{m} for preprocessor Macro
	}
	\label{fig:corpus}
	\centering
\begin{tabular}{|l|l|l|}
\hline
\textbf{Term}        & \textbf{Locations}                                                                                                  & \begin{tabular}{c}\textbf{Weight} \\ \tiny{parse\_command} \end{tabular}  \\ \hline
axis\_command        & \begin{tabular}[c]{@{}l@{}} (2,3,\textbf{i}), (6,5,\textbf{i}), (7, 7,\textbf{i}),\\ (9,12,\textbf{i}), (10,12,\textbf{i}), (14,7,\textbf{i})\end{tabular}                                    & 0.78               \\ \hline
move\_y              & (10,5,\textbf{i})                                                                                                               & 0.30                 \\ \hline
parse                & (1,6,\textbf{i}), (5,12,\textbf{i}), (7,7,\textbf{i}),                                                                    & 0.60             \\ \hline
unsupported\_command & (13,17,\textbf{m})                                                                                                             & 0.30             \\ \hline
move\_x              & (9,5,\textbf{i})                                                                                                               & 0.30                  \\ \hline
input                & \begin{tabular}[c]{@{}l@{}}(1,26,\textbf{i}), (3,7,\textbf{i}), (5,23,\textbf{i}),\\  (7,26,\textbf{i}), (11,14,\textbf{i})\end{tabular}                                    & 0.78                   \\ \hline
fail                 & (13,12,\textbf{m})                                                                                                             & 0.30             \\ \hline
coolant              & (12,5,\textbf{i})                                                                                                               & 0.30                     \\ \hline
axis                 & \begin{tabular}[c]{@{}l@{}} (2,3,\textbf{i}), (6,5,\textbf{i}), (7, 7,\textbf{i}),\\ (9,12,\textbf{i}), (10,12,\textbf{i}), (14,7,\textbf{i})\end{tabular}                          & 0.85                    \\ \hline
command              & \begin{tabular}[c]{@{}l@{}}(1,12,\textbf{i}), (2,3,\textbf{i}), (6,5,\textbf{i}), \\ (7,7,\textbf{i}), (9,12,\textbf{i}), (10,12,\textbf{i}),\\  (13,17,\textbf{i}), (14,7,\textbf{i}), (15,5,\textbf{i})\end{tabular} & 1.00                   \\ \hline
parse\_axis\_command & (7,7,\textbf{i})                                                                                                              & 0.30                    \\ \hline
null                 & (2,18,\textbf{i})                                                                                                              & 0.30                     \\ \hline
do\_command          & (15,5,\textbf{i})                                                                                                              & 0.30                    \\ \hline
unsupported          & (13,17,\textbf{i})                                                                                                             & 0.30                     \\ \hline
parse\_unit          & (5,12,\textbf{i})                                                                                                              & 0.30                 \\ \hline
move                 & (9,5,\textbf{i}), (10,5,\textbf{i})                                                                                                        & 0.48                       \\ \hline
mm                   & (4,8,\textbf{c})                                                                                                              & 0.30                      \\ \hline
inches               & (4,14,\textbf{c})                                                                                                              & 0.30                      \\ \hline
unit                 & (5,5,\textbf{i}), (5,12,\textbf{i})                                                                                                       & 0.48                      \\ \hline
mode                 & (8,5,\textbf{i}), (15,16,\textbf{i})                                                                                                      & 0.48                      \\ \hline
parse\_command       & (1,6,\textbf{i})                                                                                                               & 0.30                     \\ \hline
\end{tabular}

\end{table}

The process of creating the corpus can be quite demanding for a large project. 
Computing the SVD for a corpus is time consuming, too. 
To avoid reoccurrence of these overheads, we persist the Corpus in a serialized Java object, so that we compute the corpus and its SVD only once for a program and can re-use it for every new query.

\subsection{Extracting a corpus slice}
\label{sec:corpusslice}

To query the corpus for slices representing some feature, 
the user provides a tuple \emph{(term, similarity-threshold)}.

The term here corresponds to the string that likely represents the feature of interest. 

We envision that the domain expert will have to explore different permutations of input terms and similarity thresholds before arriving at his intended result. For every input query, a \texttt{similarity score} is computed for each document. This score measures how similar the input term is to the document; how this value is calculated depends on the IR model in use. 

We 

filter out all documents with a similarity score less than the threshold. In the resulting documents, we gather all terms related to the input term, i.e., the terms where the input term is a substring. 

For illustration, assume that we query the corpus of the example in Figure \ref{fig:code} for the term \emph{axis}. We construct the initial query vector, which has 21 entries - one for each term in the corpus Table~\ref{fig:corpus}, with a one in each position involving the query term (\emph{axis}) and zeros for the rest. This query vector is then mapped to the SVD space of the corpus in this concrete example, using LSI.
The similarity score for \emph{axis} in our case for the only document (\emph{function parse\_command}) is 1.0, which means that the document makes the cut for the next step. Next, we gather all related terms, ending up with the slice of the corpus in Table \ref{tab:axis_tdm}. 

\begin{table}[]

\caption{Relevant TDM entries for query "axis"}
\label{tab:axis_tdm}
\begin{tabular}{|l|l|}
\hline
axis\_command        & (2,3), (4,49), (6,12), (7,12), (11,7) \\ \hline
axis                 & (2,3), (4,49), (4,64), (6,12), (7,12), (11,7) \\ \hline
parse\_axis\_command & (4,64)                                                                                   \\ \hline
\end{tabular}
\end{table}

Overall, the corpus extraction\footnote{It is implemented in Kotlin.} works in two phases. In the first phase, we create the corpus as described in \ref{sec:corpuscreation}. 
In the second phase, we use a given term to retrieve the corpus slice as described in \ref{sec:corpusslice}. This phase can be repeated several times for one program, e.g., because several terms can be used to describe a single feature. The way we construct and query the corpus is modular, so that we can use different IR methods, like LSI or VSM. We have implemented our corpus extractor in a fashion that the same format can be extracted out of grep or any other text search tool, so that results from other tools can be used as input for feature extraction, as well.

\subsection{Extracting the feature}

The final step is to use the corpus slice from the previous step to reconstruct a syntactically correct code slice 
comprising the implementation of the desired feature. The feature extractor is responsible for the following:
\begin{enumerate}
	\item Use locations from the terms in the filtered corpus as initial input
	\item Extract a slice of the program that contains the statements in the locations 
	from step 1 along with their control and data-dependencies
\end{enumerate}

The feature extractor algorithm is presented in Figure~\ref{alg-extractslice}. 

\begin{figure}                     
	\footnotesize
	\begin{tabbing}
		xxx\=xxx\=xxx\=xxx\=\kill
		\textbf{Input:} $\textsc{CS}$ =  Corpus Slice as List,\\ $\textsc{CFG}$ = Control flow graph of program (LLVM representation)\\
		\textbf{Internal Variables: }$\textsc{RELEVANT\_STMTS}$\\ \\
		\textbf{Procedure Main}\\
		1.\> \textbf{for} {entry$<$$\textsc{term}$, $\textsc{lstLocation}$}$>$ $\mathit{\in}$ $\textsc{CS}$\\
		2.\>\> \textbf{for} {$\textsc{location}$ $\mathit{\in}$ $\textsc{lstLocation}$}\\
		3.\>\>\> $\textsc{node\_term}$  $\mathit{\leftarrow}$ node corresponding to $\textsc{term}$ in $\textsc{location}$\\
		4.\>\>\> $\textsc{stmt\_term}$  $\mathit{\leftarrow}$ Statement surrounding $\textsc{node\_term}$\\
		5.\>\>\> $\textsc{RELEVANT\_STMTS}$.add($\textsc{stmt\_term}$) \\
		6.\> \textbf{for} {$\textsc{STMT}$ $\mathit{\in}$ $\textsc{RELEVANT\_STMTS}$}\\
		7.\>\> ProcessStatement($\textsc{STMT}$)\\
		8.\>\> IFDS\\
		9.\> \textbf{RETURN} Program consisting of all Statements in $\textsc{RELEVANT\_STMTS}$\\
		\\ 
	
		\textbf{Procedure ProcessStatement($\textsc{Stmt}$)} \\
		10.\> \textbf{for} {$\textsc{varrefexpr}$ $\mathit{\in}$ $\textsc{STMT}$}\\
		11.\>\> $\textsc{DECLARATION}$ $\mathit{\leftarrow}$ Definition of $\textsc{varrefexpr}$ in $\textsc{CFG}$\\
		12.\>\>  $\textsc{RELEVANT\_STMTS}$.add($\textsc{DECLARATION}$)\\
		13.\> \textbf{for} {$\textsc{funccallexpr}$ $\mathit{\in}$ $\textsc{STMT}$}\\
		14.\>\> $\textsc{DEFINITION}$ $\mathit{\leftarrow}$ Definition of $\textsc{funccallexpr}$ in $\textsc{CFG}$\\
		15.\>\>  $\textsc{RELEVANT\_STMTS}$.add($\textsc{DEFINITION}$)\\
		16.\> \textbf{if} {$\textsc{STMT}$} is a return statement\\
		17.\>\> $\textsc{lstAssignedStatements}$ $\mathit{\leftarrow}$ Statements that \\
		   \>\>                          consume the return value of {$\textsc{STMT}$}\\
		18.\>\>  $\textsc{RELEVANT\_STMTS}$.add($\textsc{DEFINITION}$)\\
		19.\> \textbf{if} {$\textsc{STMT}$ $\mathit{\in}$ $\textsc{BLOCK}$}\\
		20.\>\>  $\textsc{RELEVANT\_STMTS}$.add($\textsc{CFG}$.start($\textsc{BLOCK}$))\\
		21.\>\>  $\textsc{RELEVANT\_STMTS}$.add($\textsc{CFG}$.end($\textsc{BLOCK}$))\\
		22.\> $\textsc{Stmt}$.processed  $\mathit{\leftarrow}$ true\\
		23.\> for $\textsc{STMT\_unprocessed}$ $\mathit{\in}$ $\textsc{RELEVANT\_STMTS}$ \\
		24.\>\>where $\textsc{STMT\_unprocessed}$.processed != true\\
		25.\>\>\> ProcessStatement($\textsc{STMT\_unprocessed}$)
		  	
	\end{tabbing}
	\caption{Outline of the code slicing algorithm}
	\label{alg-extractslice}
\end{figure}

\begin{enumerate}
	\item In Lines 1-5, we mark all statements that correspond to retrieved locations as relevant.
	\item In Line 7, we process each statement to find its dependencies
	\item In Lines 10-12, we mark definitions of every variable reference inside the processed statement as relevant.
	\item In Lines 13-18, we perform similar markings to handle flows between function call sites and definitions. 
	\item In lines 19-21, we mark the start and end of blocks to ensure syntactic completion of the sliced CFG.
	\item Finally, we mark the statement as processed and continue till all relevant statements are processed.  
\end{enumerate}

\begin{figure}
	\includegraphics[width=0.5\textwidth]{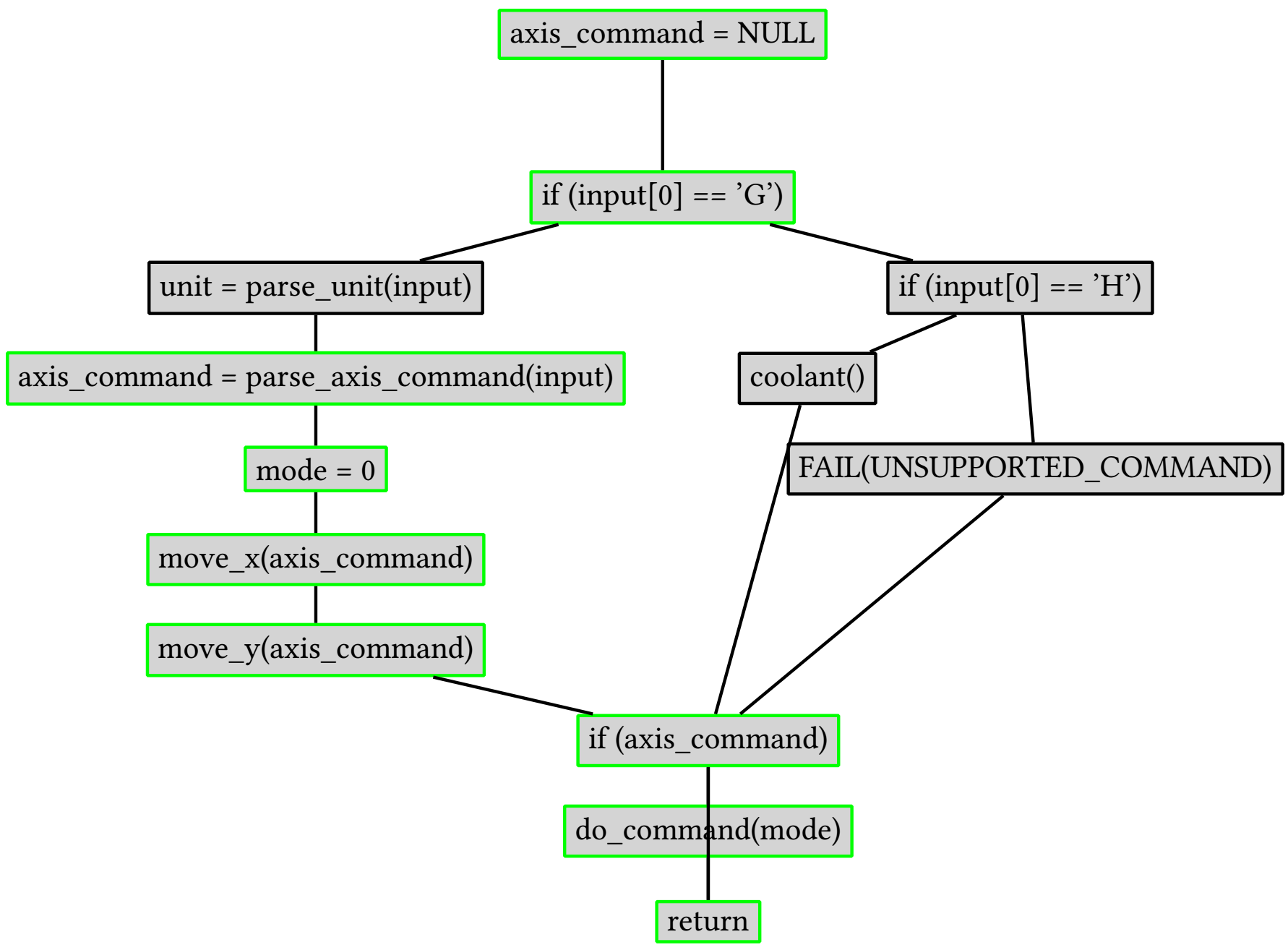}
	\caption{CFG of LLVM IR of \emph{parse\_command}}
	\label{fig:cfg}
\end{figure}

We illustrate this process on the code of Figure \ref{fig:code}, with the search term \emph{Axis}.
\begin{enumerate}
	\item Mark the statements in lines 2, 6, 7, 9, 10, 14 since they are contained in our locations retrieved from the corpus. These line numbers are the first entry of each tuple of the column \textbf{Locations} corresponding to the terms from the sliced corpus. 
	\item We add line 15 since it is in the body of the \texttt{if} in line 14 and the \texttt{if} including its condition in line 8
	\item We add line 3 since we add statements that are controlled by the \texttt{if}.
	\item We collect all block endings, where a statement is marked, i.e., the closing braces in 11,16 and 18. In addition add the function head i.e. Line 1.
\end{enumerate}
The CFG of the function \texttt{parse\_comand} in Figure~\ref{fig:code} is illustrated in Figure~\ref{fig:cfg}, 
where the relevant nodes are highlighted in green. 
The feature extractor results in the code in Fig. \ref{fig:slicedCode}.

\label{sec:extrfeature}

\begin{figure}
\begin{lstlisting}[language=c,numbers=left]
void parse_command(char* input) {
  axis_command = NULL;
  if (input[0] == 'G') {
    axis_command = 
      parse_axis_command(input);
    int mode = 0;
    move_x(axis_command);
    move_y(axis_command);
  } else if (input[0] == 'H'){
    coolant();
  } else { FAIL(UNSUPPORTED_COMMAND) }
  if (axis_command) {
    do_command(mode);
  }
}
\end{lstlisting}
\caption{Example Code- Sliced}
	\label{fig:slicedCode}
\end{figure}

The feature extractor is based on Phasar~\cite{boddenphasar} and is implemented in C++. Phasar implements the \textbf{IFDS} framework to solve inter-procedural, finite, distributive subset (IFDS) problems introduced by Reps, Horwitz and Sagiv~\cite{repspopl}. Given the IFDS framework, designers of data-flow program analyses need only define a set of flow functions, which the framework solves. In order to define a program analysis using Phasar, we need to implement the following four functions~\footnote{\url{https://github.com/secure-software-engineering/phasar/wiki/Writing-an-IFDS-analysis}}.

\begin{enumerate}
\item The analysis developer uses the \textbf{getNormalFlowFunction}
function to express what happens during intra-procedural data flows. For our implementation, we mark every intra-procedural data flow dependency of a relevant node as relevant.
\item The analysis developer uses the \textbf{getCallFlowFunction} function to express what happens when a call-site is encountered, i.e., how to handle the relationship between the actual parameters and the formal parameters of a function. For our implementation, we mark every formal parameter of an actual relevant parameter as relevant
\item The analysis designer uses the \textbf{getRetFlowFunction} function to indicate how to handle the flow from a return statement to the variable the return value is assigned to. For our implementation, we mark the variable where the value of a relevant return statement is assigned, as relevant.
\item The analysis designer uses the \textbf{getCallToRetFlowFunction} the indicate how facts flow around a call site, e.g. facts that are not modified by the callsite itself but still have to be propagated. For our implementation, we accordingly propagate all facts that still are relevant after the call statement.
\end{enumerate}

Our feature creator compiles the original source code to an SSA-compliant LLVM-IR including debug information. Single static assignment(SSA)~\cite{braunssa} is a popular attribute of any intermediate representation of code, that emphasizes that each variable be assigned only once. The advantage is that optimizations are simplified and analyzing properties of variables are facilitated. Phasar provides an SSA representation of the LLVM IR, which we use to make our analysis more efficient. We mark the initial locations returned by our corpus extractor as relevant and initiate the IFDS solver, which uses the flow function to mark the relevant control and data dependencies. We augment this with other necessary control dependencies that are not included by our slicing algorithm a.k.a unconditional jump instructions. We then use the LLVM debug information to retrieve the original source code of all marked instructions to produce the sliced version of the original source code.

\section{Experiments}
\label{sec:eval}

We validated \toolname{} by applying it to extract modules from a set of real C programs ranging between 5k and 165k lines of code. We present the setup of the experiment and its results in the following.

\subsection{Methodology}

We queried Github for C-projects. We consider a C-project any project with a significant portion of the code written in C (>80\% lines of C-code). 

We sorted all C-projects returned by the above query by the number of stars and obtained a list of candidate projects. We read descriptions from project websites searching candidates with interesting modules and identified six such projects.  For each project, one of the authors with three years of C experience manually identified modules; another author with four years of C experience independently validated them. In the case of a disagreement, both authors agreed on a solution. The manually identified modules serve as our ground truth. For each of them we defined representative search terms that capture the intention. The projects and the extracted modules are:
\begin{enumerate}
\item \textit{parson}\footnote{\url{https://github.com/kgabis/parson}}: json library, $5439$ lines of non-header code. We identified two modules to extract from \textit{parson}, one module concerned with JSON parsing and one module concerned with JSON serialization. 
\item \textit{inotify-tools}\footnote{\url{https://github.com/inotify-tools/inotify-tools}}: Collection of tools to watch for filesystem events, $4741$ lines of non-header code. We defined a module concerned with handling of internal stats, e.g., the number of file access events. 
\item  \textit{fping}\footnote{\url{https://github.com/schweikert/fping}}: A variant of ping, $2261$ lines of non-header code. For \textit{fping} we identified a module concerned with handling of internal stats, e.g., response times respectively. 
\item  \textit{silver searcher}\footnote{\url{https://github.com/ggreer/the_silver_searcher}}: Code search, $5085$ lines of non-header code. The module we extracted is responsible for filtering filenames using regular expressions. 
\item \textit{memcached}\footnote{\url{https://github.com/memcached/memcached}}: Key/value store mainly used for caching, $24262$ lines of non-header code. For \textit{memcached} we defined a stats handling module, similar to \textit{fping} and \textit{inotify-tools}, as well as a module for the handling of commands that the webserver accepts. 
\item \textit{redis}\footnote{\url{https://github.com/redis/redis}}: In memory database with a web-api, $165159$ lines of non-header code. In the case of \textit{redis} we defined a module that is concerned with the handling of the cluster-manager mode of redis' commandline interface.
 \end{enumerate}

\label{lbl:groundtruth}
We extracted each ground truth module into its own file and added corresponding header files. Hence, in some cases we had to remove or add the \textit{static} modifier to reduce or increase the visibility of functions and variables. The manually extracted modules are "able to be merged", i.e., no conflicts with the base repository and that the same unit tests (including any CI checks from the repository) pass, as they did before our refactoring. 

The external validity of our ground truth for \textbf{Silver searcher} is implicitly ensured, 
as it was defined equivalent to a code 
module on which a transpilation to Rust was performed; there has
been a pull request for the transpiled Rust version of the module 
(cf Section~\ref{sec:introduction})\footnote{\url{https://github.com/ggreer/the_silver_searcher/pull/1418}}.

\label{sec:inter_limit}
To facilitate the extraction of code from larger modules, we introduced the notion of an inter-procedural distance limit to constraint how far through call edges our tool looks up. We limited all of our experiments to two call edges, except for redis where we deactivated the inter procedural analysis.

We applied \toolname{} using the search terms (see Table~\ref{fig:evalModules}) of the ground truth modules and compared  automatically retrieved modules with the ground truth ones. For comparison, we performed a set level difference between the lines from the tool extracted output and from the ground truth. Intersecting lines are considered to be correctly extracted; 
those in the ground truth but not in the extracted modules are considered missing lines;
those in the extracted modules but not in the ground truth are considered additional.
We do not keep track of comments and exclude those from our comparison.

We perform manual analysis on each of these differing sets
 to gather insights on what sort of manual intervention,
 or future support could be necessary. In addition we use \textit{grep} to generate a baseline extraction to which we can compare.

Table~\ref{fig:evalModules} summarizes the results.

Each line lists the project, the extracted module, and the used search term in the first three columns.\footnote{As it can be seen, the query is almost similar to the expected module.} We used a similarity threshold of $0.85$ for all of the extractions. The column \textbf{lines} shows the overall number of LoCs in the module and - in parentheses - unique LoCs after excluding empty ones and comments. The meaning of  \textbf{extracted correctly}, \textbf{missing}, and \textbf{additional} is self explanatory, we display these values for both \toolname{} and the tool \textit{grep}.
All extractions complete in less than 3 minutes except for one outlier which took 34 minutes. 
Our validation consists in (a) a comparison with grep, (b) an analysis of lines that are missing from the extracted modules in comparison to the ground truth, (c) an analysis of lines that get extracted, but are not part of the ground truth, i.e. additional lines, and (d) an discussion of the impact of chosen search terms.

\begin{table*}
	\caption{Modules and Extraction Results. $\dagger$: We used different constraints for \textit{redis}, see \ref{sec:inter_limit}}
\label{fig:evalModules}	
	\begin{tabular}{|c|c|c|c||c|c||c|c||c|c||c|} \hline
		Project&Module &Search Term& Lines & \multicolumn{2}{c||}{Extracted correctly} & \multicolumn{2}{c||}{Missing} & \multicolumn{2}{c||}{Additional} & Runtime \\ 
		&&&& \toolname & \textit{grep} & \toolname & \textit{grep} & \toolname & \textit{grep}  &\\
		 \hline
		Parson&Parse    & \textit{parse}       & 400(301)    & 273    & 22            & 28 & 289    & 115 & 18       & 9 seconds     \\ 
		Parson&Serialize         & \textit{serialize}   & 360(201)    & 182 & 23                & 19   & 183  & 38 & 4        & 34 minutes      \\
		inotify&Statistics       &\textit{stats},\textit{stat}   & 270(159)    & 158 & 28               & 1 & 162     & 30 & 88      & <1 second    \\
		inotify&Statistics       & \textit{stats}   & 270(159)    & 110  & 7              & 49 & 183      & 10  & 7    &  <1 second   \\
		fping &Statistics        &\textit{stats}   & 240(140)    &  94   & 2             &   56 & 166  & 430 & 14      & 19 seconds \\
		silver searcher          & File Filter & \textit{ignore} & 260(240)    & 209  & 37             & 31 & 215    & 248 & 56       & 24 seconds\\
		memcached& Command & \textit{command,cmd} & 3150(1370) & 1023 &141 & 347 &1481 & 802&204 & 13 seconds\\ 
		memcached& Statistics & \textit{stats} & 710(485) & 365 & 8 & 120 & 512 & 2267 & 337 & 130 seconds \\ 
		redis\textsuperscript{$\dagger$} & Cluster Manager & \textit{clustermanager} & 4490(2689) & 2274 & 492 & 415 & 2501 & 10 & 39 & 11 seconds \\ \hline
	\end{tabular}
\end{table*}
\subsection{Comparison with grep}
We compare the results of our slicing approach with the regular expression search tool \textit{grep}. To the best of our knowledge, grep is the only other well known tool, with which we can apply natural language search for sub-sets of C code.
Both \textit{INFOX}~\cite{zhouforks} and \textit{CLUSTERCHANGES}~\cite{barnett-clusterchanges-2015} are not compareable, since they focus and rely on changes/changesets e.g. from forks, whereas \toolname{} takes only the current state of a project/repository as input.
For all the projects and modules we used the exact same search terms, as with \toolname{}. The results can be seen in Figure \ref{fig:evalModules}. In order to make the results comparable we removed all comments from the grep results, since we are mainly concerned with the executable/compilable source code. 
\begin{tcolorbox}
	While  \toolname{} achieved a recall for extracted module between $62.67\%$ and $99.37\%$, the alternative of using grep as a starting point to identify modules from C-programs achieved only a recall between $1.4\%$ and $18.3\%$
\end{tcolorbox}

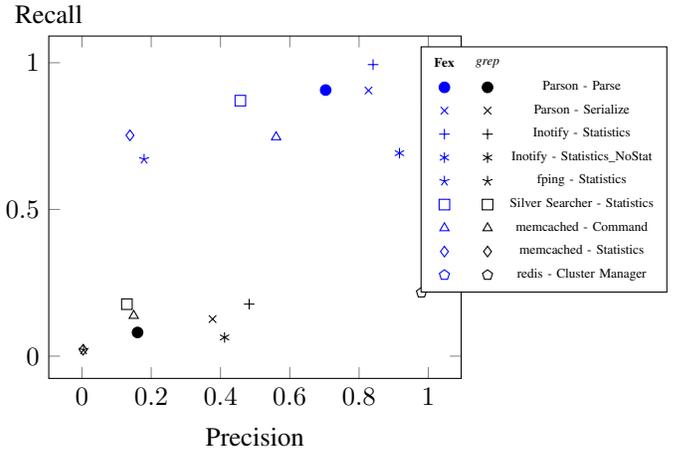
\begin{figure}
	\begin{tikzpicture}
		\begin{axis}[%
			scale=0.8,
			scatter/classes={%
				a={mark=*},
				b={mark=*,draw=black,fill=black},
				c={mark=x},
				d={mark=x,draw=black,fill=black},
				e={mark=+},
				f={mark=+,draw=black,fill=black},
				g={mark=asterisk},
				h={mark=asterisk,draw=black,fill=black},
				i={mark=star},
				j={mark=star,draw=black,fill=black},
			    k={mark=square},
  			    l={mark=square,draw=black,fill=black},
				m={mark=triangle},
				n={mark=triangle,draw=black,fill=black},
				o={mark=diamond},
				p={mark=diamond,draw=black,fill=black},
				q={mark=pentagon},
				r={mark=pentagon,draw=black,fill=black}},
			ylabel near ticks,
every axis y label/.style={at={(current axis.north west)},above=0.5mm},
			xlabel=Precision,		
			ylabel=Recall,
			legend style={font=\tiny,at={(1.5,0.97)}},
			legend columns=2,
			legend entries={
				{Parson - Parse},
				{Parson - Serialize},
				{Inotify - Statistics},
				{Inotify - Statistics\_NoStat},
				{fping - Statistics},
				{Silver Searcher - Statistics},
				{memcached - Command},
				{memcached - Statistics},
				{redis - Cluster Manager}
			},
			]
			\addlegendimage{legend image with text=\toolname}
			\addlegendentry{}
			\addlegendimage{legend image with text=\textit{grep}}
			\addlegendentry{}
			\addlegendentry{}
			\addlegendentry{{Parson - Parse}}
			\addlegendentry{}
			\addlegendentry{{Parson - Serialize}}
			\addlegendentry{}
			\addlegendentry{{Inotify - Statistics}}
			\addlegendentry{}
			\addlegendentry{{Inotify - Statistics\_NoStat}}
			\addlegendentry{}
			\addlegendentry{{fping - Statistics}}
			\addlegendentry{}
			\addlegendentry{{Silver Searcher - Statistics}}
			\addlegendentry{}
			\addlegendentry{{memcached - Command}}
			\addlegendentry{}
			\addlegendentry{{memcached - Statistics}}
			\addlegendentry{}
			\addlegendentry{{redis - Cluster Manager}}
			\addplot[scatter,only marks,draw=blue,fill=blue,scatter src=explicit symbolic]%
			table[meta=class] {
				x y class label
				0.70360824742268	0.906976744186046 a {Parson - Parse}
				0.827272727272727	0.90547263681592 c	{Parson - Serialize}
				0.840425531914894	0.993710691823899 e	{Inotify - Statistics}
				0.916666666666667	0.691823899371069 g {Inotify - Statistics\_NoStat}
				0.179389312977099	0.671428571428571 i {fping - Statistics}
				0.457330415754923	0.870833333333333 k {Silver Searcher - Statistics} 
				0.560547945205479	0.746715328467153 m {memcached - Command}
				0.138677811550152	0.752577319587629 o {memcached - Statistics} 
				0.995621716287215	0.845667534399405 q {redis - Cluster Manager} 
			};
			\addplot[scatter,only marks,scatter src=explicit symbolic]%
			table[meta=class] {
			x y class label Pos addOffset color
			0.160583941605839	0.080586080586081	b	{Parson - Parse}
			0.377049180327869	0.126373626373626	d	{Parson - Serialize} 
			0.482758620689655	0.177215189873418	f	{Inotify - Statistics}
			0.411764705882353	0.063636363636364	h	{Inotify - Statistics\_NoStat} 
			0.00462962962963	0.021276595744681	j	{fping - Statistics}
			0.129824561403509	0.177033492822967	l   {Silver Searcher - Statistics} 
			0.149522799575822	0.13782991202346	n	{memcached - Command} 
			0.003516483516484	0.021917808219178	p	{memcached - Statistics} 
			0.9800796812749		0.216358839050132	r	{redis Cluster - Manager}
		
	};	
				
			\end{axis}				
		\end{tikzpicture}
	\caption{Comparison of Precision \& Recall for \toolname{}(in blue)  and \textit{grep} (in black)}
	\label{fig:precision_recall}
\end{figure}
\subsection{Analysis of missing lines of code}
\label{sec:missing}

\begin{figure}	
	\includegraphics[scale=0.25]{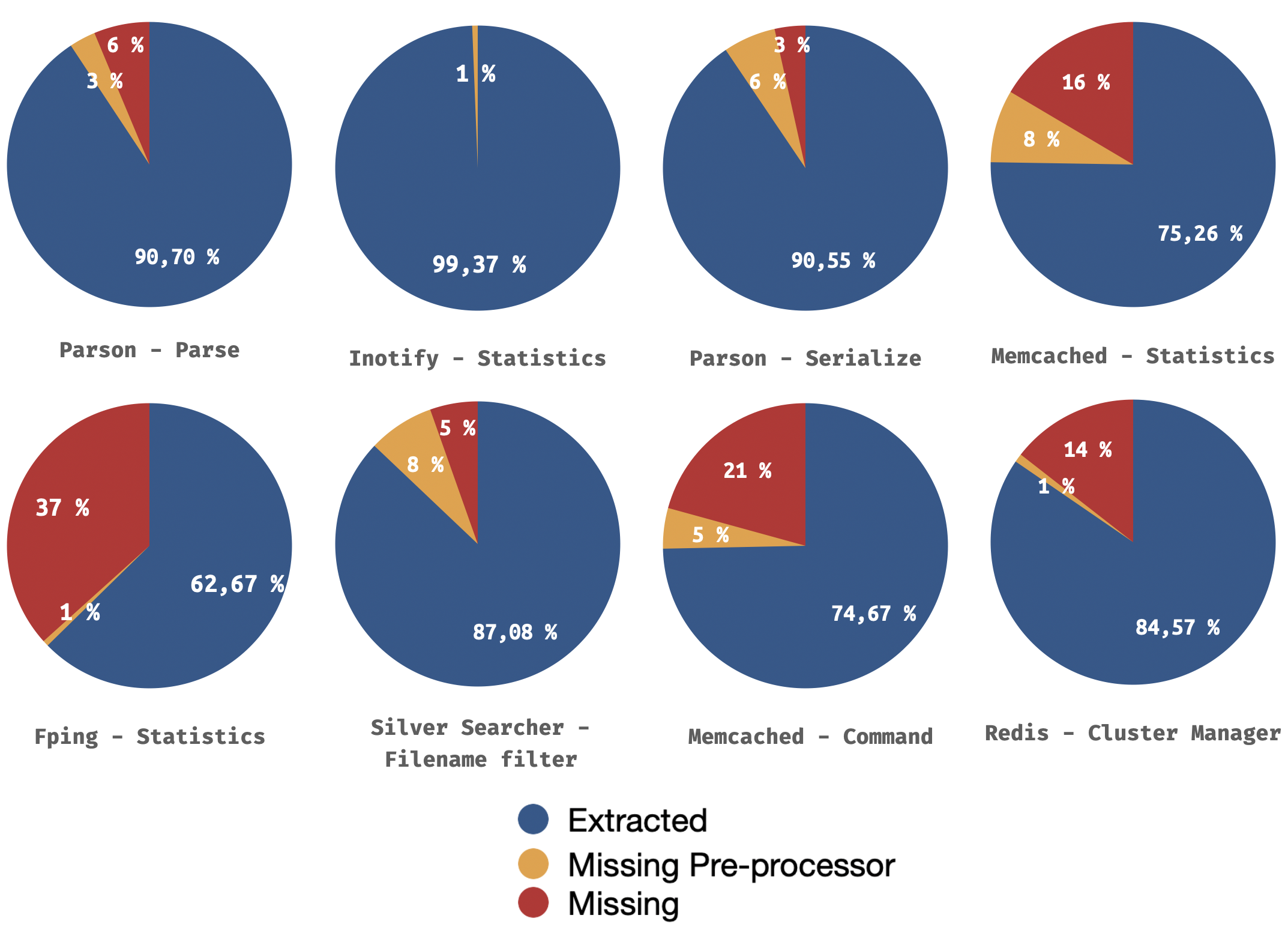}
	\caption{Missing lines relative to properly extracted ones}
	\label{fig:missingpie}
\end{figure}

Figure~\ref{fig:missingpie} elaborates on the missing lines for each of the expected modules relative to the number of lines that were correctly extracted and on the respective root causes. 

First, we highlight that all but one (\textbf{fping-statistics}) extracted modules are close to the ground truth. The portion of the correctly extracted lines (blue) constitutes roughly $75\% - 99\%$ in the respective pies. Moreover, a lot of missing lines are accounted for by preprocessor macros, which have limited support with Phasar and LLVM as they are optimization frameworks that do not intend to preserve behavior during transformations (orange parts of the pies). The \textbf{fping-statistics} module is an exception with 38\% actually missing lines.

Next, we discuss reasons for actually missing lines 
(i.e., not comments or preprocessor directives) and for additional lines.

In \textbf{fping}, 

all missing lines ($56$) result from lines that are semantically related to the module under investigation,
but do not have a data-dependence on any relevant variables. Consider e.g., line 1 in the snippet in Figure~\ref{fig:printfEx}. It prints a new line and does not refer to relevant variables (\inline|num_alive|, \inline|num_hosts|), referred to Lines 2 and 3; thus is not included in the extracted module, even if it is part of an overall code block that prints statistics.

Lines 5, 6 are not included in the extracted module either, because the print statement in line 7 is also not included. 

For \textbf{silver searcher}, our extraction misses
$8$ lines in the filename filter - this is because our tool does not readily include all lines in multi-line initializers, 
a feature that merely

needs a bit more engineering. For illustration, consider Figure~\ref{fig:arrayDecl}, where lines 2-6 are not included in the initialization. 

For \textbf{memcached-command}, we miss $347$ lines, which are in a way or another due to

our limited macro handling.

The root cause for the $120$ missing lines in the \textbf{memcached-stats} module is the same, since $40$ lines contain macro invocations.

For \textbf{redis-cluster manager}, we miss $415$ lines.

Of those, $50$ are  declarations of functions where we only extract the definitions, 
$26$ are macro usages, and the rest are due

to the wrong handling of multiline lines and struct/array declarations.

\begin{tcolorbox}
Overall, most missing lines are due to a few limitations of the current prototype: (a) handling of function/struct and array declarations and (b) handling of simple preprocessor directives, e.g. \lstinline|#include| With the exception of the \textit{fping} extraction, all of the other modules are affected by this. All of these categories are solvable with engineering effort and should not prevent the principal applicability of \toolname{}. 
\end{tcolorbox}

\subsection{Analysis of additional lines of code}
\label{sec:additional}

\begin{sloppypar}
The extractor produces 140 additional lines for the parse module of \textbf{Parson}, out of which $103$  pertain to the inclusion of methods that represent the API for JSON object modification, e.g., \inline|json_value_init_object|, \inline|json_object_resize|, etc. 
\end{sloppypar}
We did not include these methods in our ground truth since to our opinion they represent a separate module. However, during the parsing there is extensive use of functions that create and modify JSON objects and thus our extraction process includes these methods. 

\begin{tcolorbox}
With code-bases containing multiple inter-dependent modules (for parsing  and modification of JSON objects), the domain expert could apply \toolname{} repeatedly on the extracted broader parsing module to arrive at further splits, one for modification of the objects in our example. 
\end{tcolorbox}

The extracted statistics module of \textbf{fping} has 430 additional lines compared to the ground truth, $380$ of which result from the fact that the stat module is controlled by a flag called  \inline|stats_flag|. Since the search term \textit{stats} occurs in this flag, our tool marks several occurrences of this flag as relevant code, e.g., Figure \ref{fig:fpingex} lines 2,5 and 6. 
The largest chunk of erroneously extracted lines results from the snippet in lines 1-3 of Figure \ref{fig:fpingex}. Since these lines belong to the part of the code that does the main commandline parsing,  \toolname{} transitively includes most of this commandline parsing and its dependencies.  

In the \textbf{redis-cluster manager} module all $10$ additional lines stem from the command line parsing that activates the extracted module.
A large source of additional lines in our \textit{memcached} stats module extraction stems from the way stats are collected throughout the project. Figure \ref{fig:stats_api} shows such an example. Our stats module 

writes the collected stats directly into one of two global structs
They are called \lstinline|stats| or \lstinline|stats_state| respectively. 
This leads to a large number of relevant starting locations that do not represent our module, but rather usages of the module. 
This leads to the inclusion of a lot of lines that we do not consider relevant for the module.
\begin{tcolorbox}
In scenarios where the code is organized with multiple variables with similar names to the search terms, a domain expert with a little bit of knowledge about the code could rename one of the occurrences such as \inline|stats_flag| to break the chain and obtain a more representative module. 
\end{tcolorbox}

\begin{figure}[t]
	\begin{lstlisting}[language=c,numbers=left,numbersep=4pt]
		fprintf(stderr,"\n");
		fprintf(stderr," %7d targets\n", num_hosts);
		fprintf(stderr," %7d alive\n", num_alive);
		
		update_current_time();
		curr_tm = localtime((time_t*)&current_time.tv_sec);
		fprintf(stderr, "[%2.2d:%2.2d:%2.2d]\n", curr_tm->tm_hour,curr_tm->tm_min, curr_tm->tm_sec);
	\end{lstlisting}
	\caption{Example Code from \textit{fping}, that shows constructs that lead to non extracted code}
	\label{fig:printfEx}
\end{figure}
\begin{figure}[t]
	\begin{lstlisting}[language=c,numbers=left,numbersep=4pt]
...
case 's':
stats_flag = 1;
break;
// Other commandline cases
...
if (stats_flag)
print_global_stats();	
	\end{lstlisting}
	\caption{Additional lines in the stats module of \textit{fping}}
	\label{fig:fpingex}
\end{figure}
\begin{figure}[t]
	\begin{lstlisting}[language=c,numbers=left,numbersep=4pt]
const char *ignore_pattern_files[] = {
	".ignore",
	".gitignore",
	".git/info/exclude",
	".hgignore",
	NULL
};
	\end{lstlisting}
	\caption{Array Declaration in Silver Searcher}
	\label{fig:arrayDecl}
\end{figure}
\begin{figure}[t]
	\begin{lstlisting}[language=c,numbers=left,numbersep=4pt]
STATS_LOCK();
	stats_state.hash_power_level = hashpower;
	stats_state.hash_bytes = hashsize(hashpower) * sizeof(void *);
STATS_UNLOCK();
	\end{lstlisting}
	\caption{Usage of stats struct in memcached}
	\label{fig:stats_api}
\end{figure}

\subsection{Impact of the search terms}
Our approach relies on carefully selected search terms - thus, we opted for some empirical observations about the impact that the term selection may have on the resulting code. To this end, we experimented with two different sets of search terms in the case of \textit{inotify}. They are:

\begin{enumerate}
\item (\textit{stats}, \textit{stat})
\item (\textit{stats})
\end{enumerate}

In the context of files in a system, there are two concepts that may be referred to by the term \textit{stat}: The singular of the word "stats" and the linux function \texttt{stat} that returns file information. Removing the singular version (\textit{stat}) reduced the number of additional lines by $20$
 removing all calls to the linux function \textit{stat}. But, it also increased the missing lines by $48$, since there is a function called \textit{stat\_it} which is relevant for our module inside the ground truth. 

In summary, the selection of the search term impacts the results. 
This was also hinted at when we discussed additional lines. But, in a sense this is a feature and not a bug. Our tool is modular enough to accommodate for input from more sophisticated forms of search terms, like declarative queries or more sophisticated expressions as proposed in literature~\cite{query1}\cite{query2}\cite{query3}.

\begin{tcolorbox}
The goal of  \toolname{} is not to replace the human in the extraction process, rather to serve as a prequel step to do much of the heavy lifting of identifying pieces of code potentially representing a feature. It provides the engineers with means to explore features in code. Project code can at the end be cut into modules by different criteria.
\end{tcolorbox}

\section{Threats to validity and future work}
\label{sec:threats}
In this section, we discuss some threats to validity and planned future work to address those concerns.
\paragraph{Scope of informational retrieval}
Currently, we consider each function to be a document, which could be problematic in poorly modularized code-bases with large functions with many lines of code. This risk can potentially be mitigated by choosing finer-grained definitions of documents. Another aspect of informational retrieval that could be improved is to apply other indexing techniques than LSI. Such a future experiment could also serve to help us understand if additional lines resulting out of similarly sounding variables can be avoided, if more fine-grained weighting schemes to terms are considered. 
\paragraph{Choice of ground truth}
Since our ground truth modules were defined by one of the authors, there could be an element of bias involved in the selection. An ideal scenario would have been to conduct a thorough survey using domain experts, which is left for future work. However, we have tried to mitigate this issue by having another author oversee the defined ground truths and submitting pull requests to the maintainers.  The status of the pull requests as of writing this paper is as follows. \textbf{Inotify-tools} accepted the request with positive feedback.\footnote{\url{https://github.com/inotify-tools/inotify-tools/pull/129}} \textbf{Parson} closed the request without explanation.\footnote{\url{https://github.com/kgabis/parson/pull/152}} This is presumably because our retrieved modules constitute a new file each, while they explicitly want the code to be within two files; this is stated in their "About page"\footnote{\url{https://github.com/kgabis/parson}} as "\texttt{Lightweight (only 2 files)}". The rest are still not processed as of writing this paper.
\paragraph{Generality}
Since our experiments required a lot of manual effort, an evaluation with a representative set of large projects was infeasible.
We have attempted to mitigate this concern by conducting a partial analysis on two large code-bases. In addition we want to investigate the effects of the introduced inter-procedural distance limit further.
\paragraph{C preprocessor macros}
At the moment \toolname{} does not support the handling preprocessor macros, which leads to imprecision in features that use those macros. Although currently, our missing line numbers are not vastly affected by this, we are aware that preprocessor macros support is vital to our approach and We plan to add support for the handling of macros to \toolname{}, by extending \textit{SPL\textsuperscript{LIFT}} \cite{bodden-spllift-2013}.
\section{Related Work}
\label{sec:relatedwork}
To the best of our knowledge our work is the first to combine IR and slicing to extract features based on natural language input.
 
\subsection{Identifying and extracting features}

The desire to identifying features from legacy code is nothing new. CodeCarbonCopy~\cite{cccdouskous} allows developers to identify functionality to be transferred into a another part of the project. Automated Software Transplantation~\cite{barrtransplantation} is another technique that identifies functionalities from a \textit{host} based on an input code location.  Although both these approaches perform code analysis to identify dependencies of the functionality, the input is a code location, which would expect the user of the system to be well-versed with the code. In contrast, our approach allows developers to begin with a term in natural language and only expects the user to fix potential missing and additional lines later. 

Another perspective to the same problem is to look at software as a product line and manage the individual features. Approaches have been proposed to extract product lines from Software. ArgoUML-SPL~\cite{coutospl} is an open source tool that extracts Software product lines (SPL) from UML diagrams. This may be useful for identifying potential features in a large application, but in order to manifest the SPL in code, one would have to use an approach such as ours to aid with extraction of these features. Another approach that helps to organize requirements as product line requirements is CoreReq~\cite{bergercore}. 

The above mentioned approaches expect that developers to think about feature-based software development prior to the coding phase, but we have established in Section~\ref{sec:introduction} to be not always the case in reality. 

Perhaps the closest to our work from the perspective of extraction of modules are the ones that attempt to reverse engineer SPL from existing software. But4Reuse~\cite{but4reuse} is a generic approach to extract software product lines from artifacts ranging from images to source code. Although in theory their approach should work with problems such as ours especially since they have adapters for C and Java programs, source level feature extraction requires careful implementation of adapters, and these adapters cannot capture the specific semantics of a particular software, but rather are written using language-specific hints such as preprocessors. Shantnawi et al.~\cite{snatwanispl} recover SPL from already existing product variants in object-oriented code, which also relies on language-specific features to separate variants in code.

\subsection{Managing features}
Another school of thought to handling features in software is to manage them without modifying the code. Zhou et al.~\cite{zhouforks} have proposed a technique to use opportunities present in forks of branches in repository to identify features. Their approach also uses information retrieval to label commits into clusters, which provides technical leads with a visual overview of in-development features for a project. This approach and ours  
complement each other.

The C-preprocessor is often used

to identify features in a code. Peopl~\cite{behringer:icse2017} allows 

to identify features, visualize and edit them. Typechef~\cite{typechef} relies on existing variant information in preprocessor pragmas to check for type errors in variants. Malaqueis et al.~\cite{malaquias:icpc2017} harness the information present in preprocessor based C-code to provide visualization of existing variants.
\textit{Leviathan}~\cite{hofer-leviathan-2010} and the approach by St\u{a}nciulescu et al.~\cite{stanciulescu-projection-2016} provide projections for single configurations and thus enable the evolution of only one configuration. \textit{CIDE}~\cite{kaestner-cide-2012},~\cite{feigenspan2013background} provides the basis for the analysis of software product line features, for Java.

All of these approaches are helpful for abstracting, visualizing and managing existing features in source, but do not help when users want to have a starting point to even identify, where these features are inside the code-base, especially without necessarily relying on preprocessor directives.

Copy-paste-redeemed~\cite{narcrecpr} abstracts clones into modules from C source code. Linked Editing~\cite{toominlinked} does not abstract modules, but provides an opportunity to link code clones and edit them together. These approaches are complementary to ours.

\subsection{Program slicing}
Weiser~\cite{weiserslicing} introduced a framework and a process to extract program slices back in 1981, applications of which are endless~\cite{slicingapps} including ours. The second phase of our approach draws a lot of inspiration from the techniques proposed there. Infoslicer~\cite{infoslicer} is one such closely related work that proposes an algorithm to extract inter-procedural program slices. Since we use the SSA form of the LLVM's IR, it is envisionable that we could employ state of the art pointer analysis provided by SVF~\cite{svfyulei}. srcSlice~\cite{srcslice} is a highly scalable program slicing technique that is able to gather program slices for every variable in a large code-base such as the linux kernel within 15 minutes. Although such a technique would be useless for people who want to identify codes linked to a natural language term, such techniques can be plugged into our program slicing part for larger examples to avoid resource-overhead. 
\section{Conclusion}
\label{sec:conclusion}
Large software today are a collection of multiple functionalities and organizing them as individual functionalities would greatly benefit maintenance costs. Unfortunately, because many large scale software are not built with this intention prior to development phase, individual features or functionalities are largely scattered throughout the code-base. Current techniques to identify and extract such inter-dependent code bases rely on program slicing techniques that require code locations as input, something a non-expert in the code-base may not possess. To this end, we propose a novel approach that takes a natural language term as input and combines information retrieval and static analysis techniques to extract modules representing these terms. We built a prototype tool to implement this approach, evaluate it on five fairly large open source code bases and report insights based on this experiment. Our experience with the tool reveals that even though such a subjective module extraction cannot be fully automated, it can go a long way in reducing the manual effort required for such an endeavour. 
\section*{Acknowledgements}
\addcontentsline{toc}{section}{Acknowledgment}
This work was funded by the Hessian LOEWE
initiative within the Software-Factory 4.0 project.
This work has been co-funded by the Crossing SFB 119 and through the support of the National Research Center for Applied Cybersecurity ATHENE.

\bibliography{references}

\begin{thebibliography}{10}
\providecommand{\url}[1]{#1}
\csname url@samestyle\endcsname
\providecommand{\newblock}{\relax}
\providecommand{\bibinfo}[2]{#2}
\providecommand{\BIBentrySTDinterwordspacing}{\spaceskip=0pt\relax}
\providecommand{\BIBentryALTinterwordstretchfactor}{4}
\providecommand{\BIBentryALTinterwordspacing}{\spaceskip=\fontdimen2\font plus
\BIBentryALTinterwordstretchfactor\fontdimen3\font minus
  \fontdimen4\font\relax}
\providecommand{\BIBforeignlanguage}[2]{{%
\expandafter\ifx\csname l@#1\endcsname\relax
\typeout{** WARNING: IEEEtran.bst: No hyphenation pattern has been}%
\typeout{** loaded for the language `#1'. Using the pattern for}%
\typeout{** the default language instead.}%
\else
\language=\csname l@#1\endcsname
\fi
#2}}
\providecommand{\BIBdecl}{\relax}
\BIBdecl

\bibitem{Clements2001}
P.~C. Clements and L.~Northrop, ``Software product lines: Practices and
  patterns,'' ser. SEI Series in Software Engineering.\hskip 1em plus 0.5em
  minus 0.4em\relax Addison-Wesley, August 2001.

\bibitem{berger:splc2015}
\BIBentryALTinterwordspacing
T.~Berger, D.~Lettner, J.~Rubin, P.~Gr\"{u}nbacher, A.~Silva, M.~Becker,
  M.~Chechik, and K.~Czarnecki, ``What is a feature? a qualitative study of
  features in industrial software product lines,'' in \emph{Proceedings of the
  19th International Conference on Software Product Line}, ser. SPLC '15.\hskip
  1em plus 0.5em minus 0.4em\relax New York, NY, USA: Association for Computing
  Machinery, 2015, p. 16–25. [Online]. Available:
  \url{https://doi.org/10.1145/2791060.2791108}
\BIBentrySTDinterwordspacing

\bibitem{malaquias:icpc2017}
R.~{Malaquias}, M.~{Ribeiro}, R.~{Bonifácio}, E.~{Monteiro}, F.~{Medeiros},
  A.~{Garcia}, and R.~{Gheyi}, ``The discipline of preprocessor-based
  annotations - does ifdef tag n't endif matter,'' in \emph{2017 IEEE/ACM 25th
  International Conference on Program Comprehension (ICPC)}, 2017, pp.
  297--307.

\bibitem{le:vlhc2011}
D.~{Le}, E.~{Walkingshaw}, and M.~{Erwig}, ``ifdef confirmed harmful: Promoting
  understandable software variation,'' in \emph{2011 IEEE Symposium on Visual
  Languages and Human-Centric Computing (VL/HCC)}, 2011, pp. 143--150.

\bibitem{behringer:icse2017}
B.~{Behringer}, J.~{Palz}, and T.~{Berger}, ``Peopl: Projectional editing of
  product lines,'' in \emph{2017 IEEE/ACM 39th International Conference on
  Software Engineering (ICSE)}, 2017, pp. 563--574.

\bibitem{ZHAO2013105}
\BIBentryALTinterwordspacing
Y.~Zhao, ``Chapter 10 - text mining,'' in \emph{R and Data Mining}, Y.~Zhao,
  Ed.\hskip 1em plus 0.5em minus 0.4em\relax Academic Press, 2013, pp. 105 --
  122. [Online]. Available:
  \url{http://www.sciencedirect.com/science/article/pii/B9780123969637000106}
\BIBentrySTDinterwordspacing

\bibitem{deerwester-indexing-1990}
S.~Deerwester, S.~Dumais, G.~Furnas, T.~Landauer, and R.~Harshman, ``Indexing
  by latent semantic analysis.'' 1990, pp. 391--407.

\bibitem{trefethen97}
L.~N. Trefethen and D.~Bau, ``Numerical linear algebra.''\hskip 1em plus 0.5em
  minus 0.4em\relax SIAM, 1997.

\bibitem{boddenphasar}
P.~D. Schubert, B.~Hermann, and E.~Bodden, ``Phasar: An inter-procedural static
  analysis framework for c/c++,'' in \emph{Tools and Algorithms for the
  Construction and Analysis of Systems}, T.~Vojnar and L.~Zhang, Eds.\hskip 1em
  plus 0.5em minus 0.4em\relax Cham: Springer International Publishing, 2019,
  pp. 393--410.

\bibitem{repspopl}
\BIBentryALTinterwordspacing
T.~Reps, S.~Horwitz, and M.~Sagiv, ``Precise interprocedural dataflow analysis
  via graph reachability,'' in \emph{Proceedings of the 22nd ACM SIGPLAN-SIGACT
  Symposium on Principles of Programming Languages}, ser. POPL '95.\hskip 1em
  plus 0.5em minus 0.4em\relax New York, NY, USA: Association for Computing
  Machinery, 1995, p. 49–61. [Online]. Available:
  \url{https://doi.org/10.1145/199448.199462}
\BIBentrySTDinterwordspacing

\bibitem{braunssa}
M.~Braun, S.~Buchwald, S.~Hack, R.~Lei{\ss}a, C.~Mallon, and A.~Zwinkau,
  ``Simple and efficient construction of static single assignment form,'' in
  \emph{Compiler Construction}, R.~Jhala and K.~De~Bosschere, Eds.\hskip 1em
  plus 0.5em minus 0.4em\relax Berlin, Heidelberg: Springer Berlin Heidelberg,
  2013, pp. 102--122.

\bibitem{barnett-clusterchanges-2015}
M.~Barnett, C.~Bird, J.~Brunet, and S.~K. Lahiri, ``Helping developers help
  themselves: Automatic decomposition of code review changesets,'' in
  \emph{2015 IEEE/ACM 37th IEEE International Conference on Software
  Engineering}, vol.~1, 2015, pp. 134--144.

\bibitem{query1}
\BIBentryALTinterwordspacing
Z.~Shang, E.~Zgraggen, B.~Buratti, F.~Kossmann, P.~Eichmann, Y.~Chung,
  C.~Binnig, E.~Upfal, and T.~Kraska, ``Democratizing data science through
  interactive curation of ml pipelines,'' in \emph{Proceedings of the 2019
  International Conference on Management of Data}, ser. SIGMOD '19.\hskip 1em
  plus 0.5em minus 0.4em\relax New York, NY, USA: Association for Computing
  Machinery, 2019, p. 1171–1188. [Online]. Available:
  \url{https://doi.org/10.1145/3299869.3319863}
\BIBentrySTDinterwordspacing

\bibitem{query2}
\BIBentryALTinterwordspacing
Z.~Zhao, L.~De~Stefani, E.~Zgraggen, C.~Binnig, E.~Upfal, and T.~Kraska,
  ``Controlling false discoveries during interactive data exploration,'' in
  \emph{Proceedings of the 2017 ACM International Conference on Management of
  Data}, ser. SIGMOD '17.\hskip 1em plus 0.5em minus 0.4em\relax New York, NY,
  USA: Association for Computing Machinery, 2017, p. 527–540. [Online].
  Available: \url{https://doi.org/10.1145/3035918.3064019}
\BIBentrySTDinterwordspacing

\bibitem{query3}
\BIBentryALTinterwordspacing
C.~Reichenbach, Y.~Smaragdakis, and N.~Immerman, ``Pql: A purely-declarative
  java extension for parallel programming,'' in \emph{Proceedings of the 26th
  European Conference on Object-Oriented Programming}, ser. ECOOP'12.\hskip 1em
  plus 0.5em minus 0.4em\relax Berlin, Heidelberg: Springer-Verlag, 2012, p.
  53–78. [Online]. Available:
  \url{https://doi.org/10.1007/978-3-642-31057-7_4}
\BIBentrySTDinterwordspacing

\bibitem{bodden-spllift-2013}
\BIBentryALTinterwordspacing
E.~Bodden, T.~Tol\^{e}do, M.~Ribeiro, C.~Brabrand, P.~Borba, and M.~Mezini,
  ``Spl<sup>lift</sup>: Statically analyzing software product lines in minutes
  instead of years,'' \emph{SIGPLAN Not.}, vol.~48, no.~6, p. 355–364, Jun.
  2013. [Online]. Available: \url{https://doi.org/10.1145/2499370.2491976}
\BIBentrySTDinterwordspacing

\bibitem{cccdouskous}
\BIBentryALTinterwordspacing
S.~Sidiroglou-Douskos, E.~Lahtinen, A.~Eden, F.~Long, and M.~Rinard,
  ``Codecarboncopy,'' in \emph{Proceedings of the 2017 11th Joint Meeting on
  Foundations of Software Engineering}, ser. ESEC/FSE 2017.\hskip 1em plus
  0.5em minus 0.4em\relax New York, NY, USA: Association for Computing
  Machinery, 2017, p. 95–105. [Online]. Available:
  \url{https://doi.org/10.1145/3106237.3106269}
\BIBentrySTDinterwordspacing

\bibitem{barrtransplantation}
\BIBentryALTinterwordspacing
E.~T. Barr, M.~Harman, Y.~Jia, A.~Marginean, and J.~Petke, ``Automated software
  transplantation,'' in \emph{Proceedings of the 2015 International Symposium
  on Software Testing and Analysis}, ser. ISSTA 2015.\hskip 1em plus 0.5em
  minus 0.4em\relax New York, NY, USA: Association for Computing Machinery,
  2015, p. 257–269. [Online]. Available:
  \url{https://doi.org/10.1145/2771783.2771796}
\BIBentrySTDinterwordspacing

\bibitem{coutospl}
M.~V. {Couto}, M.~T. {Valente}, and E.~{Figueiredo}, ``Extracting software
  product lines: A case study using conditional compilation,'' in \emph{2011
  15th European Conference on Software Maintenance and Reengineering}, 2011,
  pp. 191--200.

\bibitem{bergercore}
\BIBentryALTinterwordspacing
I.~Reinhartz-Berger and M.~Kemelman, ``Extracting core requirements for
  software product lines,'' \emph{Requirements Engineering}, vol.~25, no.~1,
  pp. 47--65, Mar 2020. [Online]. Available:
  \url{https://doi.org/10.1007/s00766-018-0307-0}
\BIBentrySTDinterwordspacing

\bibitem{but4reuse}
\BIBentryALTinterwordspacing
J.~Martinez, T.~Ziadi, T.~F. Bissyand\'{e}, J.~Klein, and Y.~Le~Traon,
  ``Bottom-up adoption of software product lines: A generic and extensible
  approach,'' in \emph{Proceedings of the 19th International Conference on
  Software Product Line}, ser. SPLC '15.\hskip 1em plus 0.5em minus 0.4em\relax
  New York, NY, USA: Association for Computing Machinery, 2015, p. 101–110.
  [Online]. Available: \url{https://doi.org/10.1145/2791060.2791086}
\BIBentrySTDinterwordspacing

\bibitem{snatwanispl}
\BIBentryALTinterwordspacing
A.~Shatnawi, A.-D. Seriai, and H.~Sahraoui, ``Recovering software product line
  architecture of a family of object-oriented product variants,'' \emph{Journal
  of Systems and Software}, vol. 131, pp. 325 -- 346, 2017. [Online].
  Available:
  \url{http://www.sciencedirect.com/science/article/pii/S0164121216301327}
\BIBentrySTDinterwordspacing

\bibitem{zhouforks}
\BIBentryALTinterwordspacing
S.~Zhou, S.~St\u{a}nciulescu, O.~Le\ss{}enich, Y.~Xiong, A.~W\k{a}sowski, and
  C.~K\"{a}stner, ``Identifying features in forks,'' in \emph{Proceedings of
  the 40th International Conference on Software Engineering}, ser. ICSE
  '18.\hskip 1em plus 0.5em minus 0.4em\relax New York, NY, USA: Association
  for Computing Machinery, 2018, p. 105–116. [Online]. Available:
  \url{https://doi.org/10.1145/3180155.3180205}
\BIBentrySTDinterwordspacing

\bibitem{typechef}
\BIBentryALTinterwordspacing
A.~Kenner, C.~K\"{a}stner, S.~Haase, and T.~Leich, ``Typechef: Toward type
  checking ifdef variability in c,'' in \emph{Proceedings of the 2nd
  International Workshop on Feature-Oriented Software Development}, ser. FOSD
  '10.\hskip 1em plus 0.5em minus 0.4em\relax New York, NY, USA: Association
  for Computing Machinery, 2010, p. 25–32. [Online]. Available:
  \url{https://doi.org/10.1145/1868688.1868693}
\BIBentrySTDinterwordspacing

\bibitem{hofer-leviathan-2010}
\BIBentryALTinterwordspacing
W.~Hofer, C.~Elsner, F.~Blendinger, W.~Schr\"{o}der-Preikschat, and D.~Lohmann,
  ``Toolchain-independent variant management with the leviathan filesystem,''
  in \emph{Proceedings of the 2nd International Workshop on Feature-Oriented
  Software Development}, ser. FOSD '10.\hskip 1em plus 0.5em minus 0.4em\relax
  New York, NY, USA: Association for Computing Machinery, 2010, p. 18–24.
  [Online]. Available: \url{https://doi.org/10.1145/1868688.1868692}
\BIBentrySTDinterwordspacing

\bibitem{stanciulescu-projection-2016}
S.~Stănciulescu, T.~Berger, E.~Walkingshaw, and A.~Wąsowski, ``Concepts,
  operations, and feasibility of a projection-based variation control system,''
  in \emph{2016 IEEE International Conference on Software Maintenance and
  Evolution (ICSME)}, 2016, pp. 323--333.

\bibitem{kaestner-cide-2012}
\BIBentryALTinterwordspacing
C.~K\"{a}stner, S.~Apel, T.~Th\"{u}m, and G.~Saake, ``Type checking
  annotation-based product lines,'' \emph{ACM Trans. Softw. Eng. Methodol.},
  vol.~21, no.~3, Jul. 2012. [Online]. Available:
  \url{https://doi.org/10.1145/2211616.2211617}
\BIBentrySTDinterwordspacing

\bibitem{feigenspan2013background}
J.~Feigenspan, C.~K{\"a}stner, S.~Apel, J.~Liebig, M.~Schulze, R.~Dachselt,
  M.~Papendieck, T.~Leich, and G.~Saake, ``Do background colors improve program
  comprehension in the\# ifdef hell?'' \emph{Empirical Software Engineering},
  vol.~18, no.~4, pp. 699--745, 2013.

\bibitem{narcrecpr}
K.~{Narasimhan} and C.~{Reichenbach}, ``Copy and paste redeemed (t),'' in
  \emph{2015 30th IEEE/ACM International Conference on Automated Software
  Engineering (ASE)}, 2015, pp. 630--640.

\bibitem{toominlinked}
M.~{Toomim}, A.~{Begel}, and S.~L. {Graham}, ``Managing duplicated code with
  linked editing,'' in \emph{2004 IEEE Symposium on Visual Languages - Human
  Centric Computing}, 2004, pp. 173--180.

\bibitem{weiserslicing}
M.~Weiser, ``Program slicing,'' in \emph{Proceedings of the 5th International
  Conference on Software Engineering}, ser. ICSE '81.\hskip 1em plus 0.5em
  minus 0.4em\relax IEEE Press, 1981, p. 439–449.

\bibitem{slicingapps}
L.~Du and P.~Cai, ``A survey on applications of program slicing,'' in
  \emph{Soft Computing in Information Communication Technology}, J.~Luo,
  Ed.\hskip 1em plus 0.5em minus 0.4em\relax Berlin, Heidelberg: Springer
  Berlin Heidelberg, 2012, pp. 215--220.

\bibitem{infoslicer}
Y.~{Sun}, Y.~{Zhang}, and J.~{Qian}, ``Program slicing method of llvm ir based
  on information-flow analysis,'' in \emph{2019 International Conference on
  Cyber-Enabled Distributed Computing and Knowledge Discovery (CyberC)}, 2019,
  pp. 383--390.

\bibitem{svfyulei}
\BIBentryALTinterwordspacing
Y.~Sui and J.~Xue, ``Svf: Interprocedural static value-flow analysis in llvm,''
  in \emph{Proceedings of the 25th International Conference on Compiler
  Construction}, ser. CC 2016.\hskip 1em plus 0.5em minus 0.4em\relax New York,
  NY, USA: Association for Computing Machinery, 2016, p. 265–266. [Online].
  Available: \url{https://doi.org/10.1145/2892208.2892235}
\BIBentrySTDinterwordspacing

\bibitem{srcslice}
C.~D. {Newman}, T.~{Sage}, M.~L. {Collard}, H.~W. {Alomari}, and J.~I.
  {Maletic}, ``srcslice: A tool for efficient static forward slicing,'' in
  \emph{2016 IEEE/ACM 38th International Conference on Software Engineering
  Companion (ICSE-C)}, 2016, pp. 621--624.

\end{thebibliography}
\end{document}